Title: Adiabatic theory of off-center vibronic polarons: Local dynamics on a planar lattice
Author: M. Georgiev (Institute of Solid State Physics, Bulgarian Academy of Sciences, 1784 Sofia, Bulgaria)
Comments: 14 pages pdf format
Subj-class: cond-mat

We review basic theoretical concepts and developments regarding the local and itinerant properties of off-center vibronic polarons in crystals. These include the electron self-trapping and the local rotation of the species on a square planar lattice. Phase transitions within the gas of vibronic small polarons are also discussed.

1. Introduction

The problem of itinerant small- or large- polarons in ionic, covalent or mixed compound materials has gained considerable interest for some time with regard to their transport properties. In particular itinerant polarons in layered perovskites have been considered as possible carriers of superconducting currents in what is now known as the high-$T_c$ superconductivity.[1-3] Because of the expected interrelation of high-$T_c$ superconductivity with ferroelectricity, itinerant vibronic polarons have also been mentioned as the possible candidates.[4]

In a material with strong electron-phonon coupling, the motion of an electron across the lattice is accompanied by the formation of a configurational distortion which moves along with. The mechanism of electron-distortion interplay is the distortion creating an attractive potential for the electron in which the electron becomes self-trapped. But self-trapping is not at all spontaneous and takes only place for electrons which have penetrated through a (self-trapping) barrier. The self-trapped entity composed of the electron & the associated distortion is termed polaron and under certain conditions it may become itinerant or remain bound otherwise. Depending on the spatial extension of the lattice distortion, the polaron may be small, intermediate, or large. The lattice distortion around a moving electron develops fully if the extra charge moves slowly with respect to the frequency of the coupled lattice vibration, or in energy terms, if the free-electron bandwidth is small with respect to the characteristic energy of the lattice distortion. In this case the electron motion is antiadiabatic, as it cannot follow the motion of the nuclei. In contrast, the distortion is incomplete and the electron faster than the nuclei is nearly free at the other adiabatic extreme.

If the electron-vibrational coupling is vibronic in character, it lowers the local symmetry around the moving charge. The vibronic polaron is therefore associated with a broken-symmetry lattice distortion. Depending on the nature of the symmetry-breaking interaction, we distinguish between Jahn-Teller and Pseudo-Jahn-Teller itinerant polarons.

This paper is aimed at setting out as clearly as possible the conditions under which an itinerant vibronic polaron forms. For that purpose we first discuss the self-trapping mechanism by deriving the inherent rate equation with phonon-coupled self-trapping coefficient γ and self-trapped lifetime τ. We do this by using a model self-trapping electron potential which applies to a simplified distortion pattern. Our insight into the problem deepens gradually as we define a general vibronic Hamiltonian covering both the local and itinerant behavior. For a broken-symmetry distortion, the local behavior is dominated by an off-center displacement and a roundabout rotation. These featurs occur in the 1st- and 3rd-order of the electron-phonon mixing interaction, respectively.

## 2. Rate equation

To begin with, we postulate a simple first-order rate equation to cover the local formation and decay of self-trapped polarons. Let n(t) be the free-electron density, m(t) be the self-trapped polaron (STP) density, the total density of electrons, free and self-trapped being $n_0 = n + m$. We also set N to be the total density of self-trapping sites. The STP density m(t) obtains from

$dm/dt = \gamma N n - m/\tau$,

or using $n = n_0 - m$,

$dm/dt + (\gamma N + 1/\tau)m = \gamma N n_0$.

The general solution reads:

$m(t) = \exp(-[\gamma N + 1/\tau] t) \int_0^t \gamma N n_0 \exp([\gamma N + 1/\tau] t') dt'$

$= m_{st}\{1 - \exp(-[\gamma N + 1/\tau] t)\}$

where

$m_{st} = n_0/(1 + 1/\gamma N \tau)$

is the steady-state STP density. There are two extreme cases: $m_{st} = n_0$ ($\gamma N \tau \gg 1$) and $m_{st} \ll n_0$ ($\gamma N \tau \ll 1$). We define the local antiadiabaticity in itinerancy as $t \gg 1/(\gamma N + 1/\tau)$, $m = m_{st}$ and the local adiabaticity in itinerancy as $t \ll 1/(\gamma N + 1/\tau)$, $m(t) = n_0 \gamma N t \ll m_{st}$.

We also define the aggregate time for forming a STP at time t:

$dt/d(m/n_0) = (1/m_{st})[\gamma N + 1/\tau]^{-1} \exp([\gamma N + 1/\tau] t)$

$= (1/\gamma N) \exp([\gamma N + 1/\tau] t) = \tau_{st} \exp([1/\tau_t + 1/\tau] t)$

Here $\tau_{st} = 1/\gamma N$ is the electron self-trapping time.

## 3. Electron self-trapping coefficient

The general phonon-coupled bimolecular trapping coefficient γ will be derived in a way similar to Pekar's [5]:

$$\gamma = (\gamma_d^{-1} + \gamma_b^{-1})^{-1} \quad (1)$$

where $\gamma_d$ is the trapping coefficient of a diffusion (migration) controlled process and $\gamma_b$ is its counterpart of a binding limited process. Following Pekar we have

$$\gamma_d = 4\pi D_h / \{ \int_0^{1/r_0} exp[U(r)/k_BT] \, d(1/r) \}^{-1} \quad (2)$$

where $D_h = \mu_h (k_BT/e)$ is the diffusion coefficient and $m_e$ is the free electron mobility, e, $k_B$ and T being the electronic charge, Boltzmann's constant and absolute temperature, respectively,

$$\gamma_b = \mathfrak{R}_R V_0 \quad (3)$$

where $\mathfrak{R}_R$ is the trapping rate, $V_0 = (4/3)\pi r_0^3$ is the trapping volume. U(r) is the electrostatic potential at the self-trapped site.

The electrostatic potential in self-trapping builds up self-consistently as the electron "digs out its own potential well". Nevertheless, it can be modelled as a pre-existing attractive potential U(r) which is Coulombic at long range and square-well at short range.[6] We choose a simplified form:

$$U(r) = -U_0 [1-\theta(r-r_0)] - (e^2/\kappa r)\theta(r-r_0) \quad (4)$$

r is the position vector, $r_0$ is the trapping radius, θ(r) is Heaviside's step function, $U_0 = e^2/\kappa r_0$ is the square-well potential, κ is an appropriate dielectric constant. Inserting in (2) we get

$$\int_0^{1/r_0} exp[U(r) / k_BT] \, d(1/r) = (1/a)[1- exp(-a/r_0)] \sim 1/a, \; a = e^2/\kappa k_BT \quad (5)$$

if the trapping radius $r_0$ is assumed small to replace the upper integration limit by infinity. Introducing $\gamma_c$, the migration-limited trapping coefficient of an attractive Coulomb center, we rewrite (2) to read:

$$\gamma_d \sim \gamma_c = 4\pi e\mu/\kappa \quad (6)$$

which is the migration-controlled rate component.

The trapping rate $\mathfrak{R}_R$ is phonon-coupled. Coupling to a local mode Q at a self-trapped site will arise from the electrostatic potential (4) modulated by the mode coordinate Q:

$$U(r,Q) = -(GQ+U_0)[1-\theta(r-r_0-Q)] - (e^2/\kappa r)\theta(r-r_0-Q) \quad (7)$$

Equation (7) gives rise to a linear-coupling coefficient

$$b(r) \equiv \partial U(r,Q)/\partial Q^3 \big|_{Q=0} = -G[1-\theta(r-r_0)] + [(e^2/\kappa r)-U_0]\delta(r-r_0) \tag{8}$$

where the second term is vanishing if the electrostatic potential (4) is to be continious at the $r = r_0$ boundary.

The Hamiltonian of electron, mode, and their coupling term is

$$H = H_e + H_{ph} + H_{e-ph} \tag{9}$$

where

$$H_e = \mathbf{p}_e^2/2m_e + U(r,0) \tag{10}$$

$$H_{ph} = \tfrac{1}{2}[\mathbf{P}^2/M + KQ^2] \tag{11}$$

$$H_{e-ph} = b(r)Q \tag{12}$$

where $K = M\omega^2$ is the mode force constant. In the adiabatic approximation the mode kinetic energy $\mathbf{P}^2/2M$ is discarded and Schroedinger's equation is solved with the adiabatic Hamiltonian $H_{AD} \equiv H - \mathbf{P}^2/2M$. We choose a basis of two eigenstates of the electronic Hamiltonian $H_e + H_{e-ph}$: a free electron state $|\mathbf{k}\rangle$ with energy $E_\mathbf{k}$ and a bound electron state $|b\rangle$ with energy $E_b$, the latter state being regarded as a single-electron trapped state. The adiabatic eigenvalues are then found to read

$$E_\pm(Q) = \tfrac{1}{2}\{H_{bb} + H_{\mathbf{kk}} \pm [(H_{bb} - H_{\mathbf{kk}})^2 + 4H_{b\mathbf{k}}]^{1/2}\} \tag{13}$$

where the matrix elements are

$$H_{bb} \equiv \langle b|H_{AD}|b\rangle = \tfrac{1}{2}KQ^2 + b_{bb}Q + E_b = \tfrac{1}{2}K(Q-Q_b)^2 - (E_{LR} - W + U_0),$$

$$(Q_b = -b_{bb}/K,\ E_{LR} = \tfrac{1}{2}KQ_b^2 = b_{bb}^2/2K)$$

$$H_{\mathbf{kk}} = \langle \mathbf{k}|H_{AD}|\mathbf{k}\rangle = \tfrac{1}{2}KQ^2 + b_{\mathbf{kk}}Q + E_\mathbf{k}$$

$$H_{b\mathbf{k}} = H_{\mathbf{k}b}^\dagger = \langle b|H_{AD}|\mathbf{k}\rangle = b_{b\mathbf{k}}Q \tag{14}$$

with the coupling constants

$$b_{bb} = \langle b|b(r)|b\rangle = -G$$

$$b_{\mathbf{kk}} = \langle \mathbf{k}|b(r)|\mathbf{k}\rangle = 0$$

$$b_{b\mathbf{k}} = \langle b|b(r)|\mathbf{k}\rangle = -G\langle b^3|\mathbf{k}\rangle_{in}, \tag{15}$$

etc. The bound-state minimum at $Q = Q_b$ in $H_{bb}$ is stabilized by the lattice relaxation energy $E_{LR} = G^2/2K$ and the square-well energy $U_0$ and destabilized by the bound hole kinetic energy $E_b$, roughly equal to the free hole half-bandwidth W. Inasmuch as the overlap of a free electron state with the bound state is infinitesimally small the phonon-coupled transition rate $\Re_{Rk}$ from $|k\rangle$ to $|b\rangle$ will be computed as the rate of a nonadiabatic electron transfer:[7]

$$\Re_{Rk}(T) = (2/\eta^2\omega)(\pi k_B T / E_{LR})^{\frac{1}{2}} \sinh(\eta\omega/2k_B T) |H_{bk}(Q_{bk})|^2 \exp(-E_{bk}/k_B T) \quad (16)$$

Here

$$E_{bk} = \tfrac{1}{2}KQ_{bk}^2 + b_{kk}Q_{bk} = (W - U_0 - E_k)^2 / 4E_{LR}$$

is the crossover barrier of the parabolae $H_{bb}$ and $H_{kk}$ relative to the minimum energy of the latter at $E_k$, which crossover occurs at

$$Q_{bk} = (E_b - E_k)/(b_{kk} - b_{bb}) = (W - U_0 - E_k) / G$$

We next sum up over the energy in the sea of free electron states to obtain

$$\Re_R = (2/\eta^2\omega)(\pi k_B T / E_{LR})^{\frac{1}{2}} \sinh(\eta\omega/2k_B T) \sum_{E_k} |H_{bk}(Q_{bk})|^2 \exp(-E_{bk}/k_B T) \quad (17)$$

Inasmuch as we have

$$H_{bk}(Q_{bk}) = b_{bk}Q_{bk} = -(W - U_0 - E_k)\langle b|k\rangle_{in} \quad (18)$$

inserting in (17) and converting summation to integration we get

$$\Re_R = (2/\eta^2\omega)(\pi k_B T / E_{LR})^{\frac{1}{2}} \sinh(\eta\omega/2k_B T) \int |H_{bk}(Q_{bk})|^2 \exp(-E_{bk}/k_B T) \sigma(E_k) dE_k$$

$$= (2/\eta^2\omega)(\pi k_B T/E_{LR})^{\frac{1}{2}} \sinh(\eta\omega/2k_B T) \times$$

$$\int \langle b|k\rangle\langle k|b\rangle_{in} (W-U_0-E)^2 \exp(-[W-U_0-E]^2 / 4E_{LR}k_B T)\sigma(E) dE \quad (19)$$

where $\sigma(E)$ is the electron density of states. For a lower-energy electron $\langle b|k\rangle\langle k|b\rangle_{in}$ may be factorized out of the integral.

To derive a binding rate (19) in 2D, we set $\sigma(E) = N_s / W$ for a planar lattice, where $N_s$ is the number of planar sites and W is the free-electron half-bandwidth.[8] We get

$$\gamma_b = (2/\eta^2\omega)(\pi k_B T / E_{LR})^{\frac{1}{2}} \sinh(\eta\omega/2k_B T)V_0 \times$$

$$\langle\langle b|k\rangle\langle k|b\rangle_{in}\rangle \int (W-U_0-E)^2 \exp(-[W-U_0-E]^2/4E_{LR}k_B T) \sigma(E)dE$$

$$= [2V_0\pi^{\frac{1}{2}}N_s E_{LR}(k_B T)^2/W\eta^2\omega]\langle\langle b|k\rangle\langle k|b\rangle_{in}\rangle \sinh(\eta\omega/2k_B T) \times$$

$$\{u\exp(-u^2) + (\tfrac{1}{2}\pi^{\frac{1}{2}})[1 - F(u)]\} \quad (20)$$

$$F(u) = (2/\pi^{\frac{1}{2}}) \int_0^u \exp(-x^2)dx, \quad u^2 = (W-U_0)^2/4E_{LR}k_B T.$$

$u^2$ is related to the self-trapping barrier which obtains as $E_{b0}= \frac{1}{2}KQ_{b0}^2 = (W+U_0)^2/4E_{LR}$. At $u \gg 1$ rate (20) is proportional to $exp(-u^2) = \exp(-E_{bo}/k_BT)$. $E_{st} = E_{LR} - (W-U_0)$ is the self-trapping well depth.

Considerations so far have dealt with the charge carrier coupling to a symmetry-retaining mode along its planar coordinates $Q_S$. It leads to the formation of Holstein's small polaron. To extend the model so as to cover the effect of the dual coupling to a symmetry-breaking mode $Q_A$ as well, we consider a simple case where the electron is self-trapped by $Q_S$ in excited state at a molecular site and then its energy is lowered further through coupling to $Q_A$ in ground state at that site. Now, the adiabatic potential energy surface (APES), single-well along $Q_S$, will turn double-well along $Q_A$ if the $Q_A$- coupling is sufficiently strong. If $Q_A$ couples to the charge carrier through ground state-to-excited state mixing (cf. section 4.1), then a vibronic polaron will occur. We see just how the sequence of steps as described above would end up with the formation of a vibronic polaron. Consequently, the physical implications of the dual coupling may prove far reaching.

## 4. General vibronic Hamiltonian

### 4.1. Site representation

The traditional two-band vibronic Hamiltonian is now extended so as to incorporate higher-order terms in the electron-vibrational mode interaction part, as follows:[4,9,10]

$$H = \sum_{(ij)mn} t_{ijmn} a_{im}^\dagger a_{jn} + \sum_{im} \varepsilon_{im} a_{im}^\dagger a_{im} + \frac{1}{2}\sum_{ijmn} [K_{ijmn} Q_{ijmn}^2 + \mathbf{P}_{ijmn}^2/M_{ijmn}] +$$

$$\sum_{ijmn} [G_{ijmn} Q_{ijmn} + \sum_{m'n'} C_{ijmnm'n'} Q_{ijmn} Q_{ijm'n'} + \sum_{m''n''} D_{ijmnm'n'm''n''} Q_{ijmn} Q_{ijm'n'} Q_{ijm''n''}] a_{im}^\dagger a_{jn}$$

$$+ \frac{1}{2}\sum_{ijmn} w_{(ij^3mn)} a_{im}^\dagger a_{jn}^\dagger a_{jm} a_{in} \quad (21)$$

Here m, m', m", n, n', n" = $g,u$ are band labels, i, j, … are site labels, $a_{ig}^\dagger (a_{ig})$ creates (annihilates) a $g$-band electron at site i, while $a_{ju}^\dagger (a_{lu})$ does so for a $u$-band electron at site j. Both bands are assumed narrow and closely spaced. Each band is composed of states of a given parity though $g$- and $u$- are of the opposite parities. (ij) label neighbouring sites and $t_{ijmn}$ is the hopping integral between them. $Q_{ijmn}$, $P_{ijmn}$ and $M_{ijmn}$ are the coordinates, momenta and reduced masses of the coupled vibration.

We distinguish between intra- (i = j) or inter- (i ≠ j) site vibrations coupled to the electronic system through the constants $G_{ijmn}$, $C_{ijmnm'n'}$, and $D_{ijmnm'n'm''n''}$. While the intersite mode $Q_{ijmm}$ at i ≠ j is assumed even parity (gerade), as it couples to the itinerant intraband $g$- or $u$- electrons, the intrasite mode $Q_{iimn}$ at m ≠ n is odd parity (ungerade), as it promotes $g$-band to $u$-band mixing transitions. Inasmuch as Hamiltonian (1) should conserve parity, the quadratic mixing constants $C_{iimnm'n'}$ at m,m' ≠ n,n' should be all vanishing. The second- and third-order terms are generally assumed inferior to the first-order ones.

$w_{(ij^3mn)}$ are two-particle interaction constants. In particular a m-electron at site j couples to the electric dipole $p_{mn}$ induced through m-n mixing at another electron site i:

$$w_{(ij^3mn)} = U_{ij}\delta_{mn} - \mathbf{p}_{imn}\cdot(e_j\mathbf{R}_{ij}/\kappa R_{ij}^3)(1 - \delta_{mn}) \qquad (22)$$

$\mathbf{p}_{imn}$ is the m-n mixing dipole, $U_{ij} = e_i e_j /\kappa R_{ij}$ is the intersite Coulomb repulsion.

### 4.2. Adiabatic exclusion

Excluding the lattice in the adiabatic approximation leads to a renormalized electronic Hamiltonian:[9]

$$H = \sum_{(ij)mn} T_{ijmn}\, a_{im}^\dagger a_{in} + \sum_{im} E_{im}\, a_{im}^\dagger a_{im} + \tfrac{1}{2}\sum_{ijmn} W_{(ij^3mn)}\, a_{im}^\dagger a_{jn}^\dagger a_{jm} a_{in} \qquad (23)$$

with

$T_{llgu} = \tfrac{1}{2}\varepsilon_{gu}\, [1- (\varepsilon_{gu}/4\varepsilon_{JTgu})] / 2\,sinh\,(\xi_{llgu}^2)$

$\xi_{llgu}^2 = (2\varepsilon_{Jtgu}/\eta\omega_{llgu})[1 - (\varepsilon_{gu}/4\varepsilon_{JTgu})^2]^{3/2}$

$E_{im} = \varepsilon_{im} - \varepsilon_{JTgu}\,[1+ (\varepsilon_{gu}/4\varepsilon_{JTgu})^2] + \tfrac{1}{2}\,\eta\omega_{iigu}$

$\omega_{iigu} = \omega_{iigu}\,[1- (\varepsilon_{gu}/4\varepsilon_{JTgu})^2]^{\tfrac{1}{2}}$

$W_{(lj^3gu)} = -\,\mathbf{p}_{ljgu}\cdot(e_j\mathbf{R}_{lj}/\kappa R_{lj}^3)$

$$\mathbf{p}_{lgu} = \mathbf{p}_{lgu}\,[1- (\varepsilon_{gu}/4\varepsilon_{JTgu})^2] \qquad (24)$$

locally along the mixing-mode coordinate $Q_{llgu}$ at $Q_{ijmm} = 0$. Similar equations hold good for quantities renormalized through coupling to the intersite modes $Q_{ijmm}$ at $Q_{llgu} = 0$ though now the Coulomb energy enters in lieu of the dipole-dipole term. Here $\varepsilon_{JTmn} = G_{ijmn}^2/2K_{ijmn}$ are Jahn-Teller energies, $\varepsilon_{gu} = |\varepsilon_g - \varepsilon_u|$. Because of the local mixing, the renormalized hopping term $T_{ijmn}$ splits into two components: an on-site tunneling part $T_{iigu}$ and an inter-site hopping part $T_{ijmm}$. We get a pairing energy

$$U_{ijlgu} = \tfrac{1}{2}\,\alpha_{lgu}\,(e_i\mathbf{R}_{il}/\kappa R_{il}^3 + e_j\mathbf{R}_{jl}/\kappa R_{jl}^3)^2 \qquad (25)$$

which is of the monopole--induced-dipole type, with

$$\alpha_{lgu} = \mathbf{p}_{lgu}^2/E_{gu} = a_{lgu}(\varepsilon_{gu}/E_{gu})[1- (\varepsilon_{gu}/4\varepsilon_{JT})^2], \qquad (26)$$

the vibronic polarizability, where $E_{gu} = 2T_{llgu}$, and

$$a_{lgu} = p_{lgu}^2/\varepsilon_{gu} \qquad (27)$$

is the electronic polarizability. $\alpha_{mn}$ and $a_{mn}$ correspond to the low- and high-frequency limits of polarizability.[11] Because of the polaron-band narrowing $\alpha_{mn}/a_{mn} > 1$ for $4\varepsilon_{JTmn} > \varepsilon_{mn}$.

### 4.3. Polaron pairing

Another choice of interaction constant leads to pairing into bipolarons or Cooper pairs:

$$w_{(ij^3mn)} = U_{ij}\delta_{mn} + (2\mathbf{p}_{mn}\cdot\mathbf{p}_{nm}/\kappa R_{ij}^3)(1- \delta_{mn}) \qquad (28)$$

which renormalizes to

$$W_{(ij^3mn)} = U_{ij}\delta_{mn} + (2\mathbf{p}_{mn}\cdot\mathbf{p}_{nm}/\kappa R_{ij}^3)(1- \delta_{mn}) \qquad (29)$$

It gives rise to a pairing energy of the Van der Waals type

$$U_{lmn} = (T_{llmn}/2)(\alpha_{lmn}/\kappa R_{ij}^3)^2 \qquad (30)$$

The intraband charge carriers are off-centered polarons, *viz.* electrons coupled to low-symmetry clusters. They pair forming bipolarons with a *translational mass*:

$$m_t = \eta^2/2t_b d_b^2 = \eta^2 U_b/4t_p^2 d_b^2 \qquad (31)$$

where $T_p = T_{ijmm}$, $U_b = U_{VdW}$, while $d_b$ is the bipolaron *hopping distance*. These bipolarons will Bose-condense at:[9]

$$\Theta_{cb} = 3.31\eta^2 n_b^{2/3} / k_B m_t^{D/3} m_l^{1-D/3} \qquad (32)$$

where D is the dimensionality of conduction, e.g. D = 2 for an in-plane conductivity, $m_l$ is the *local mass* of transversal off-center tunneling on the planes:

$$m_l = \eta^2/2T_{llgu}d_{ll}^2 \qquad (33)$$

However if the binding energy is much too low, Cooper pairing may precede with a condensation temperature:[12]

$$\Theta_{cp} = 1.14(\eta\omega/k_B) \exp\{-N / 2E_{JT} [N_g(0)N_u(0)]^{\frac{1}{2}}\} \qquad (34)$$

where N is the number of unit cells, while $N_g(0)$ and $N_u(0)$ are the densities of states at the Fermi level for *g* and *u*.

### 4.4. Band representation

Hamiltonian (21) is closely related to ones describing ferroelectricity in crystals regarded as a cooperative pseudo-Jahn-Teller phenomenon.[13] Expanding (21) into band waves we get:

$$H = \sum_{m\mathbf{k}}E_m(\mathbf{k}) a_{m\mathbf{k}}^\dagger a_{m\mathbf{k}} + \sum_{\mathbf{q}v} \eta\omega_v(\mathbf{q}) b_{\mathbf{q}v}^\dagger b_{\mathbf{q}v} +$$

$$\sum_{\mathbf{k,q},v} G_{\mathbf{q}v}(\mathbf{k})[\eta\omega_v(\mathbf{q})/2K_v(\mathbf{q})N]^{\frac{1}{2}}(b_{\mathbf{q}v}^\dagger b_{-\mathbf{q}v}+h.c.)(a_{m\mathbf{k}}^\dagger a_{n\mathbf{q-k}}+a_{n\mathbf{k}}^\dagger a_{m\mathbf{q-k}}) \quad (35)$$

where v is the phonon-mode polarization. A renormalised phonon frequency obtains from (35) at the central Q = 0 configuration:

$$\omega_{ren\mathbf{q}}^2 = \omega_\mathbf{q}^2 [1 + \Pi_\mathbf{q}(\omega_\mathbf{q})] \quad (36)$$

with a second-order polarisation operator

$$\Pi_\mathbf{q}(\omega_\mathbf{q}) = 4E_{JT}\Omega(2\pi)^{-3}\int d\mathbf{k} \, [E_g(\mathbf{k})-E_u(\mathbf{k-q})][n_g(\mathbf{k-q})-n_u(\mathbf{k})] /$$

$$\{[E_g(\mathbf{k})-E_u(\mathbf{k-q})]^2 - (\eta\omega_\mathbf{q})^2\} \quad (37)$$

where $\Omega$ is the unit-cell volume, $n_m(\mathbf{k})$ is the number of particles in the m-th band. $\Pi_\mathbf{q}$ being always negative, it may become large enough along a wave vector $\mathbf{q}$ to turn $\omega_\mathbf{q}$ imaginary rendering the central configuration unstable against $\mathbf{q}$. There will be a phase transition at $\omega_\mathbf{q} = 0$.

In the limit of narrow electronic bands:

$$n_m(\mathbf{k}) = \exp(-E_m(\mathbf{k})/2k_B T) / 2\cosh(E_m(\mathbf{k})/2k_B T)$$

$$\Pi_\mathbf{q}(0) = -(4E_{JT}/E_{gu})\tanh(E_{gu}/4k_B T) \quad (38)$$

From (37) we see that only the *g*-holes will exert a destabilizing effect on the paraphase, while *u*-holes will tend to stabilize it. The narrow-band model is physically equivalent to a system of noninteracting Cu-O molecules, each with a double-well potential

$$E_\pm(Q) = \tfrac{1}{2}\{KQ^2 \pm [(2GQ)^2+E_{gu}^2]^{\frac{1}{2}}\} \quad (39)$$

along $Q = Q_{\|gu}$. The free energy of weakly-interacting vibronic dipoles will therefore be

$$\Im = -2k_B\Theta\log\{\cosh[(4G^2Q^2+E_{gu}^2)^{\frac{1}{2}}/4k_B T]\} + \tfrac{1}{2}KQ^2$$

$$\pm (2p_e^2/\kappa R^3)[4G^2Q^2/(4G^2Q^2+E_{gu}^2)] \quad (40)$$

where the signs (±) are for antiferroelectric or ferroelectric alignments, respectively. The average equilibrium configurations <Q> of the lower-symmetry phase will be found from $\partial\Im/\partial Q = 0$:

$$1 = \tanh[(4G^2Q^2+E_{gu}^2)^{\frac{1}{2}}/4k_B\Theta]\times[4E_{JT}/(4G^2Q^2+E_{gu}^2)^{\frac{1}{2}}] \pm$$

$$(2p_e^2/\kappa R^3 E_{JT})[4E_{JT}E_{gu}/(4G^2Q^2+E_{gu}^2)]^2$$

Now the Curie temperature obtains from <Q> = 0 resulting in

$$\Theta_{cf} = (E_{gu}/4k_B) / \tanh^{-1}[(E_{gu}/4E_{JT}) \pm 4(2a_e/\kappa R^3)] \tag{41}$$

Because of the *tan⁻¹z* term in the denominator which effectively exceeds 1 only at z ~1, equations (32) and (41) can yield transition temperatures matching each other at reasonable values of the entry parameters for the ferroelectric alignment only.

### 5. Local rotation

We begin by pointing out that no principal distinction can be made between a small polaron and an atom at a polaron site in a lattice, as far as the local configuration is concerned. In a related example, the local electronic behavior is dominated by the coupling to a $T_{1u}$ ungerade vibration in a cubic lattice. The respective intrasite vibronic Hamiltonian of the preceding Section 4 simplifies, as follows:[14,15]

$$H = \sum_l \varepsilon_l (|\alpha\rangle\langle\beta_l| |\beta_l\rangle\langle\alpha|) + \tfrac{1}{2}\sum_l [P_l^2/M_l + M_l\omega_l^2 Q_l^2] + \sum_{\alpha\beta} [\sum_i b_i Q_I + \sum_{ij} c_{ij} Q_i Q_j +$$
$$\sum_{ijk} d_{ijk} Q_i Q_j Q_k]_{\alpha\beta} (|\alpha\rangle\langle\beta_l| + |\beta_l\rangle\langle\alpha|) \tag{42}$$

where $Q_i$ (i = x,y,z) are the $T_{1u}$ coordinates, $\alpha = a_{1g}$, $\beta = t_{1u}$, and b, c (= 0), d are the mixing constants. The adiabatic energy eigenvalues in the $\{|\alpha\rangle, |\beta_i\rangle\}$ basis are:

$$E_\pm(Q_l) = \tfrac{1}{2}M\omega^2 Q^2 \pm \{(bQ)^2 + 2b[(d_c-d_b)\sum_i Q_i^4 + d_b Q^4] + (\varepsilon_{\alpha\beta}/2)^2\}^{\tfrac{1}{2}} \tag{43}$$

for $d_{ijj} = d_b$, $d_{iii} = d_c$, $d_{ijk} = 0$ otherwise, $b_i = b$ (i,j,k=x,y,z); $\varepsilon_{\alpha\beta} = |\varepsilon_\alpha - \varepsilon_\beta|$.

For sufficiently strong electron-mode coupling, the system is displaced off-center at $Q_0$ which obtains by minimizing at d = 0:

$$Q_0 = \sqrt{\{(2E_{JT}/K)[1 - (\varepsilon_{\alpha\beta}/4\varepsilon_{JT})^2]\}}, \tag{44}$$

where $\varepsilon_{JT} = b^2/2M\omega^2$ is the Jahn-Teller energy.[15] The resulting off-center polaron can rotate upon the off-center sphere of radius $Q_0$ which rotation is governed by $E_\pm(Q_l)$ at $Q \approx Q_0$:

$$E_\pm(Q_0) = \pm[(d_c-d_b)\sum_i Q_i^4 + d_b Q_0^4](M\omega^2/b) + \varepsilon_{JT}[(1\pm 2) - (\varepsilon_{\alpha\beta}/4\varepsilon_{JT})^2] \tag{45}$$

The rotational equation is $H_{vib} u(Q) = E_{vib} u(Q)$.

In *3-D* the rotational Hamiltonian reads

$$H_{vib}(3D) = -(\eta^2/2I)\Delta_{\theta,\varphi} \pm (I\omega_{renII}^2/4)\{[(\cos\varphi\sin\theta)^4 +$$
$$(\sin\varphi\sin\theta)^4 + (\cos\theta)^4] + d_b/(d_c-d_b)\} + \varepsilon_{JT}[(1\pm 2) - (\varepsilon_{\alpha\beta}/4\varepsilon_{JT})^2] \tag{46}$$

$I = MQ_0^2$ is the inertial moment, while the renormalized frequencies are

$$\omega_{renII} = \omega_{bare}[4(d_b-d_c)/b]^{\tfrac{1}{2}} Q_0 = \omega_{renI}[8\varepsilon_{JT}(d_b-d_c)/bK]^{\tfrac{1}{2}}$$

$$\omega_{renI} = \omega_{bare} \, [1- (\varepsilon_{\alpha\beta} / 4\varepsilon_{JT})^2]^{½} \qquad (47)$$

The eigenfunctions and eigenvalues of $H_{vib}(3D)$ are so far unknown.

For a 2-D rotation along an off-center circle in the equatorial plane

$$H_{vib}(2D) = -(\eta^2 / 2I)(\partial^2/\partial\varphi^2) + \varepsilon_{JT} \, [(1\pm2) - (\varepsilon_{\alpha\beta}/4\varepsilon_{JT})^2] \pm$$
$$(I\omega_{renII}^2/4)\{-¼(3 + \cos(4\varphi)) + d_b/(d_b-d_c)\} \qquad (48)$$

where the 2-D APES is:

$$E_\pm(\varphi) = \pm(I\omega_{renII}^2/4)\{-¼(3 + \cos(4\varphi)) + d_b/(d_b-d_c)\} +$$
$$\varepsilon_{JT} \, [(1\pm2) - (\varepsilon_{\alpha\beta} / 4\varepsilon_{JT})^2] \qquad (49)$$

The reorientational barrier on $E_\pm(\varphi)$ amounts to

$$\varepsilon_{BII} = I\omega_{renII}^2/8, \qquad (50)$$

while the adiabatic energy splitting between $E_-(\varphi)$ and $E_+(\varphi)$ at $\varphi=0$ is

$$E_{12} = 4\{\varepsilon_{BII} \, [d_c/(d_b-d_c)] + \varepsilon_{JT}\}. \qquad (51)$$

### 5.1. Rotational eigenstates and bands

The eigenfunctions and eigenvalues of (46) in 3-D are so far unknown. The *2-D* rotational eigenvalue equation is Mathieu's equation:

$$-(\eta^2/2I)(\partial^2 Y/\partial\varphi^2) + 2B_\pm cos(4\varphi)Y + (C_\pm - E)Y = 0, \qquad (52)$$

$$B_\pm = \mu(\varepsilon_{BII}/4), \; C_\pm = \pm½\varepsilon_{BII} \, [(3d_c+d_b)/(d_b-d_c)] + \varepsilon_{JT} \, [(1\pm2) - (\varepsilon_{\alpha\beta}/4\varepsilon_{JT})^2]$$

with $Y=Y(\varphi)$, Mathieu's equation reading traditionally:[16]

$$d^2Y/dz^2 + [a - 2q \cos(2z)]Y = 0, \qquad (52')$$

$$q = 2(B_\pm I/\eta^2) = \mu(2\varepsilon_B/\eta\omega_{renII})^2$$

now $Y=Y(z)$ and $z=2\varphi$. There are two types of periodic eigenstates $ce_m(z,q)$ and $se_m(z,q)$ with associated eigenvalues $a_m(q)$ and $b_m(q)$ yielding rotational energies in two branches of allowed energy bands at $q < 0$ and $q > 0$, respectively:

$$E_{am,bm}(q) = (\eta^2/2I)a_m(q), b_m(q) + C_\pm. \qquad (53)$$

### 5.2. Reorientational rate

We define a two-site reorientational rate along $E_+(\varphi)$ with regard to the intrasite rotation:[7,10]

$$\Re_{12}(T) = \eta^{-1} \sum_n W_n(E_n) \, exp(-E_n / k_B T) \Delta E_{n/m=0} \sum^\infty exp(-\eta^2 [m^2 + c_m(q)] / 2I k_B T) \quad (54)$$

where $\Delta E_n = E_{n+1} - E_n$ while the sums extend over the allowed energy levels and either of $a_m(q)$, $b_m(q) = m^2 + c_m(q)$. The configurational tunneling probability $W_n(E_n)$ in $\Re_{12}(T)$ is defined by:[7]

$$W_n(E_n) = 4\pi^2 |V_{fi}|^2 \sigma_i(E_n) \sigma_f(E_n)$$

$$V_{fi} = (-\eta^2/2I) [Y_f^*(dY_i/d\varphi) - Y_i(dY_f^*/d\varphi)]_{\varphi=\varphi_c} \quad (\varphi_c = \pi/4, \, z_c = \pi/2)$$

$$\sigma(E_{a/b,n}) = dn / dE_{a/b,n} = (2I/\eta^2)(dn/da_n, b_n) \quad (55)$$

Considering linear combinations by Mathieu's functions promoting intraband transitions along $E_+(\varphi)$, we construct band states:

$$ce_{nm}(z,q) = (1-m) \, ce_{n-1}(z,q) + m \, ce_n(z,q) \quad (n \text{ odd})$$

$$se_{nm}(z,q) = (1-m) \, se_{n-1}(z,q) + m \, se_n(z,q) \quad (n \text{ even}), \quad (56)$$

where $k = \pi m$ is the wave number ($m \leq 1$), with intraband eigenvalues

$$E_{nm}(q) = (1-m)E_{n-1} + mE_n = (\eta^2/2I)a_{nm}(q),$$

$$a_{nm}(q) = (1-m)a_{n-1}(q) + ma_n(q) \quad (n \text{ odd})$$

$$E_{nm}(q) = (1-m)E_{n-1} + mE_n = (\eta^2/2I)b_{nm}(q),$$

$$b_{nm}(q) = (1-m)b_{n-1}(q) + mb_n(q) \quad (n \text{ even}) \quad (57)$$

and extend these eigenvalues so as to cover the negative m too:

$$E_{mn} = E_{n-1} + m(E_n - E_{n-1}) \; (0<m<1), \quad E_{mn} = E_n + m(E_n - E_{n-1}) \; (-1<m<0) \quad (58)$$

The normalized intraband transition probabilities read:

$$W_{Ln}(E_{nm}) = 64N \times \begin{cases} (1-m)ce_{n-1}(z,q)[m \, dce_n(z,q)/dz]^{3^2}{}_{z=\frac{1}{2}\pi}(dm/da_{nm})^2 \\ (1-m)se_{n-1}(z,q)[m \, dse_n(z,q)/dz]^{3^2}{}_{z=\frac{1}{2}\pi}(dm/db_{nm})^2 \end{cases}$$

$$N^{-1} = (128/30) \sum_{n=1}^\infty \times \begin{cases} |ce_{n-1}(z,q)[dce_n(z,q)/dz]|^2{}_{z=\frac{1}{2}\pi}(a_n - a_{n-1})^{-2} \\ |se_{n-1}(z,q)[dse_n(z,q)/dz]^3|^2{}_{z=\frac{1}{2}\pi}(b_n - b_{n-1})^{-2} \end{cases}$$

$$(59)$$

Inserting into the rate formula, we get:

$$\Re_{12}(T) = (16\eta/\pi I)(N/Z) \sum_{n=1}^{\infty} (2/g_n)\{-(1/g_n^2)[1+(6/g_n)+(12/g_n^2)]exp(-E_n/k_B T) +$$

$$[2-(6/g_n)+(13/g_n^2)-(18/g_n^3)+(12/g_n^4)]exp(-E_{n-1}/k_B T)\} \times$$

$$[a_n(q)-a_{n-1}(q)]^{-1} {}^3|ce_{n-1}(z,q)[dce_n(z,q)/dz]^3|^2_{z=\frac{1}{2}\pi}$$
$$\times \{$$
$$[b_n(q)-b_{n-1}(q)]^{-1} |se_{n-1}(z,q)[dse_n(z,q)/dz]|{}^3{}^2_{z=\frac{1}{2}\pi}$$

$$Z = \sum_{n=1}^{+\infty} (2g_n)[1-exp(-g_n)]exp(-E_{n-1}/k_B T) \qquad (60)$$

where $g_n = (E_n-E_{n-1})/k_B T$. A zero-point rate obtains reading:

$$\Re_{12}(0) = (32\eta/\pi I)N[a_1(q)-a_0(q)]^{-1} |{}^3ce_0(z,q)[dce_1(z,q)/dz]^3|^2_{z=\frac{1}{2}\pi} \qquad (61)$$

which accounts for the contribution of the lowest allowed band.

## 6. Conclusion

In the foregoing, we addressed selected items of the vibronic-polaron terminology, such as self trapping which builds up the species, either localized or itinerant, the general vibronic Hamiltonian describing the local and itinerant behavior, and the local rotation of off-center species occuring due to the broken symmetry. All these are important aspects of polaron physics introduced and studied separately in the localized and itinerant limits because of the mathematical complexities involved.